\documentclass{elsart}
\usepackage{graphics}

\newdimen\captwidth
\newdimen\normalfigwidth
\normalfigwidth=\captwidth                
\newskip\captskip
\def\etal{{\it{}et~al.}}
\def\normalfigure#1{\hbox to\textwidth{%
       \hfil\resizebox{\normalfigwidth}{!}{\includegraphics{#1}}\hfil}}
\def\capt#1{\refstepcounter{figure}\vskip\captskip\hbox to \textwidth{%
       \hfil\vbox{\hsize=\captwidth\renewcommand{\baselinestretch}{1}\small
       {\sc Figure \thefigure}\quad#1}\hfil}}
\def\eref#1{(\protect\ref{#1})}
\def\fref#1{\protect\ref{#1}}
\def\d{{\rm d}}

\begin{document}

\journal{Paleobiology}

\begin{frontmatter}
\title{Decline in extinction rates and\\
scale invariance in the fossil record}
\author[SFI]{M. E. J. Newman}
and
\author[SFI,Smithsonian]{Gunther J. Eble}
\address[SFI]{Santa Fe Institute, 1399 Hyde Park Road, Santa Fe, NM 87501.
  U.S.A.}
\address[Smithsonian]{Department of Paleobiology, Smithsonian
  Institution,\\
  Washington, DC 20560.  U.S.A.}
\begin{abstract}
  We show that the decline in the extinction rate during the Phanerozoic
  can be accurately parameterized by a logarithmic fit to the cumulative
  total extinction.  This implies that extinction intensity is falling off
  approximately as the reciprocal of time.  We demonstrate that this
  observation alone is sufficient to explain the existence of the proposed
  power-law forms in the distribution of the sizes of extinction events and
  in the power spectrum of Phanerozoic extinction, results which previously
  have been explained by appealing to self-organized critical theories of
  evolutionary dynamics.
\end{abstract}
\end{frontmatter}

\section{Introduction}
\begin{figure}[t]
\normalfigure{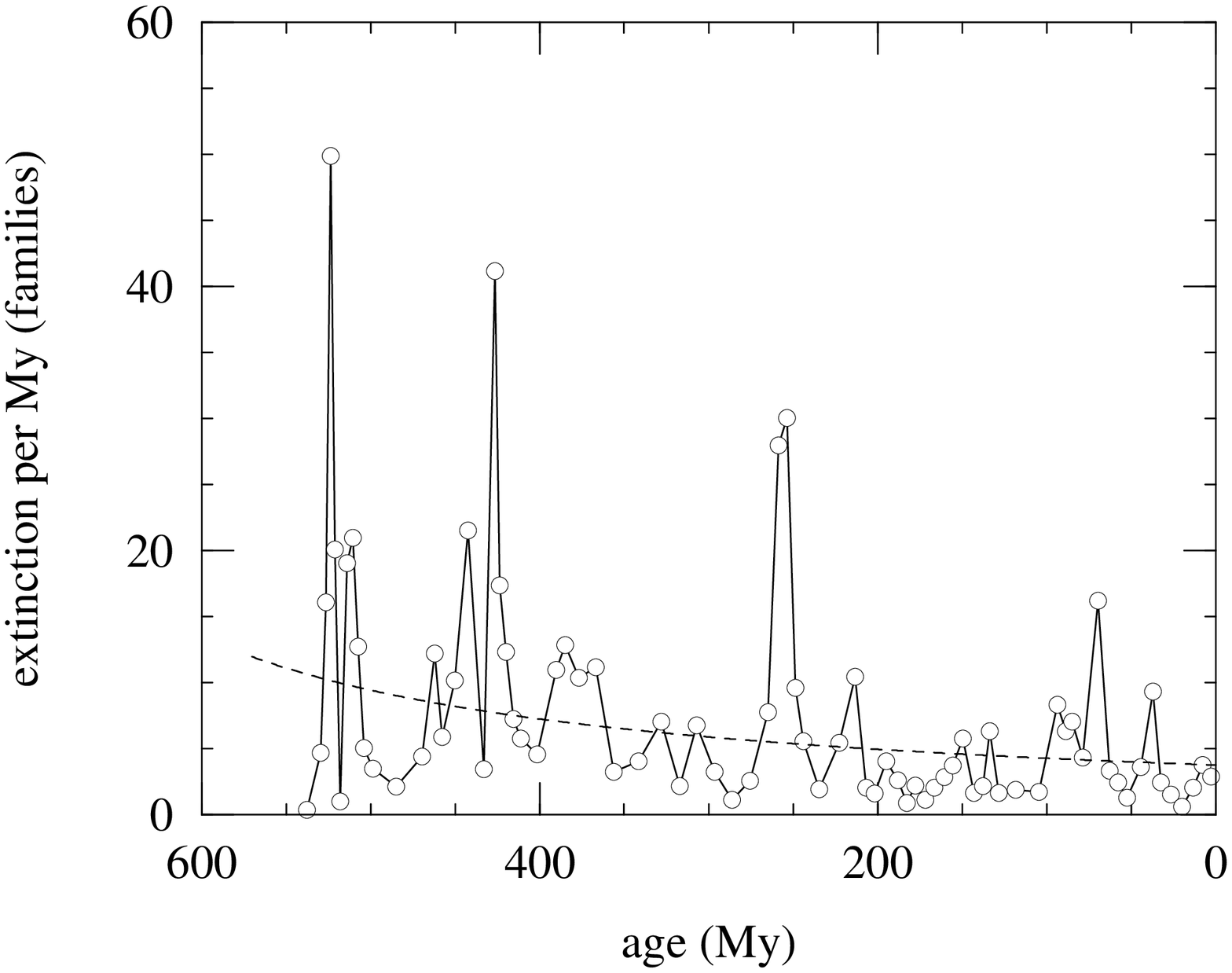}
\capt{The number of families of known marine organisms becoming extinct per
  unit time in each stratigraphic stage as a function of time during the
  Phanerozoic.  The dotted line is the average decline in extinction rate
  calculated from Equation~\eref{decay}.  The data are from the compilation
  of Sepkoski~(1992).}
\label{decline}
\end{figure}

It has been widely accepted for some time now that the mean rate of
extinction, either of genera or of families, in the known fossil record,
appears to decline throughout the Phanerozoic (Raup and Sepkoski~1982,
Sepkoski~1984, 1991, 1993, 1996, Van~Valen~1984, Flessa and Jablonski~1985,
Gilinsky and Bambach~1987, Raup~1988, Gilinsky~1994).  This decline is
illustrated in Figure~\fref{decline}, where we show the number of marine
families becoming extinct per unit time in each stage since the beginning
of the Cambrian, using data drawn from the compilation by
J.~J.~Sepkoski,~Jr.\ (Sepkoski~1992).  The decline is only an average
trend.  Certainly there are deviations from it; two of the largest
extinctions in the plot, for example, occur in the second half of the data
set.  Overall, however, the drop in extinction is clear in the figure.

The same decline is visible in other extinction metrics as well.  For
example, it has become increasingly common in recent years to quote
extinction figures in terms of the percentage of taxa becoming extinct in a
given interval, i.e.,~the ratio of the number of taxa becoming extinct to
the number in existence.  This metric shows an even more dramatic decline
than the simple extinction rate shown in Figure~\fref{decline}, because of
the steady increase in diversity throughout the Phanerozoic (Sepkoski~1993,
1996, Benton~1995).

The origin of the decline in extinction rate has been much debated.  Some
authors have argued that it is a real biological phenomenon arising from a
slow average increase in the fitness of species (Raup and Sepkoski~1982),
from taxon sorting (Sepkoski~1984, Gilinsky~1994), from changes in
taxonomic structure (Flessa and Jablonski~1985), or from taxon aging
effects (Boyajian~1986).  Others, by contrast, have suggested that it may
be an artifact stemming from sampling biases (Pease~1992).  In this paper
however, we are concerned not with the causes but with the implications of
extinction decline.

In recent years there has been a considerable amount of interest in the
so-called self-organized critical theories of evolution and extinction.
Theories of this type contend that patterns of connectivity in ecosystems
have a tendency to evolve to a ``critical'' state in which small
perturbations can trigger disturbances of arbitrary magnitude.  Starting
with Kauffman and Johnsen~(1991), a number of authors have suggested that
the fossil record of extinction shows evidence of processes of this kind
(Kauffman~1993, Sneppen~\etal~1995, Sol\'e and Bascompte~1996, Bak and
Paczuski~1996), principally because of the existence of apparently
scale-free distributions in the fossil data, particularly in the sizes of
extinction events and in the power spectrum of extinction.  Kauffman~(1993)
for instance has suggested that a histogram of the number of families
becoming extinction per stage has a power-law form.  Distributions
following power laws are often taken as indicative of critical behavior,
although there are also many other ways in which power laws can be
generated.  Sol\'e~\etal~(1997) have suggested that the power spectrum of
fossil extinction intensity---the square of the Fourier transform---also
appears to follow a power law.

In this paper we wish to make a number of observations.  First, we note
that the decline in extinction rate during the Phanerozoic can be quite
accurately parameterized if we consider the {\em cumulative\/} total
extinction, i.e.,~the total number of species which become extinct over a
period of time, as the length of that period is increased.  We show that
this quantity can be well fit by a logarithmic function.  We then use this
parameterization to calculate the expected distribution of the sizes of
extinction events in the fossil record, and show that the results are in
precise agreement with the power-law form found by other authors.  Thus
there is no need to invoke self-organized critical theories to explain this
form.  Finally, we calculate the power spectrum of our declining extinction
rate and show that it too takes a power-law form similar to that found by
Sol\'e~\etal~(1997).

\section{Parameterization of extinction decline}
Sepkoski's database of marine families gives us an estimate of the total
number of families becoming extinct in each of 77 stages dating from the
Vendian--Cambrian boundary at about 544~Ma.  (The time scale used in this
paper is that of Harland~\etal~(1990), updated with corrections kindly
supplied by J.~J.~Sepkoski,~Jr. and D.~H.~Erwin.)  Let us denote by $x(t)$
the number of families becoming extinct in the stage beginning at time $t$.
Then the {\bf cumulative total extinction} $c(t)$ at time $t$ is defined to
be the total number of families which became extinct during {\em or prior
  to\/} this stage.  Mathematically we can write this as
\begin{equation}
c(t) = \int_0^t x(t')\>\d t',
\label{cumulative}
\end{equation}
where time is measured from an origin at the start of the data set
(i.e.,~at approximately 544~Ma).  This cumulative extinction metric was
first proposed and studied by Sibani~\etal~(1995, 1998).  It is useful for
two reasons.  First, it is much less susceptible to noise in the data than
raw extinction metrics.  As we will see, the plot of $c(t)$ as a function
of $t$ is quite smooth, making it much easier to interpret than most other
metrics.  (This immunity to noise is a general characteristic of integrated
metrics, and the use of such metrics is common in many branches of science
and engineering where noise levels are high enough to be problematic.)
Second, the cumulative extinction does not depend on how time is
partitioned in the data set.  In the present study our data are divided
into stratigraphic stages, but the numerical value of $c(t)$ would be the
same if we were to choose any other partitioning.  To see this, we need
only notice that the number of taxa becoming extinct before a certain time
is not dependent on how we divide those taxa up.

\begin{figure}[t]
\normalfigure{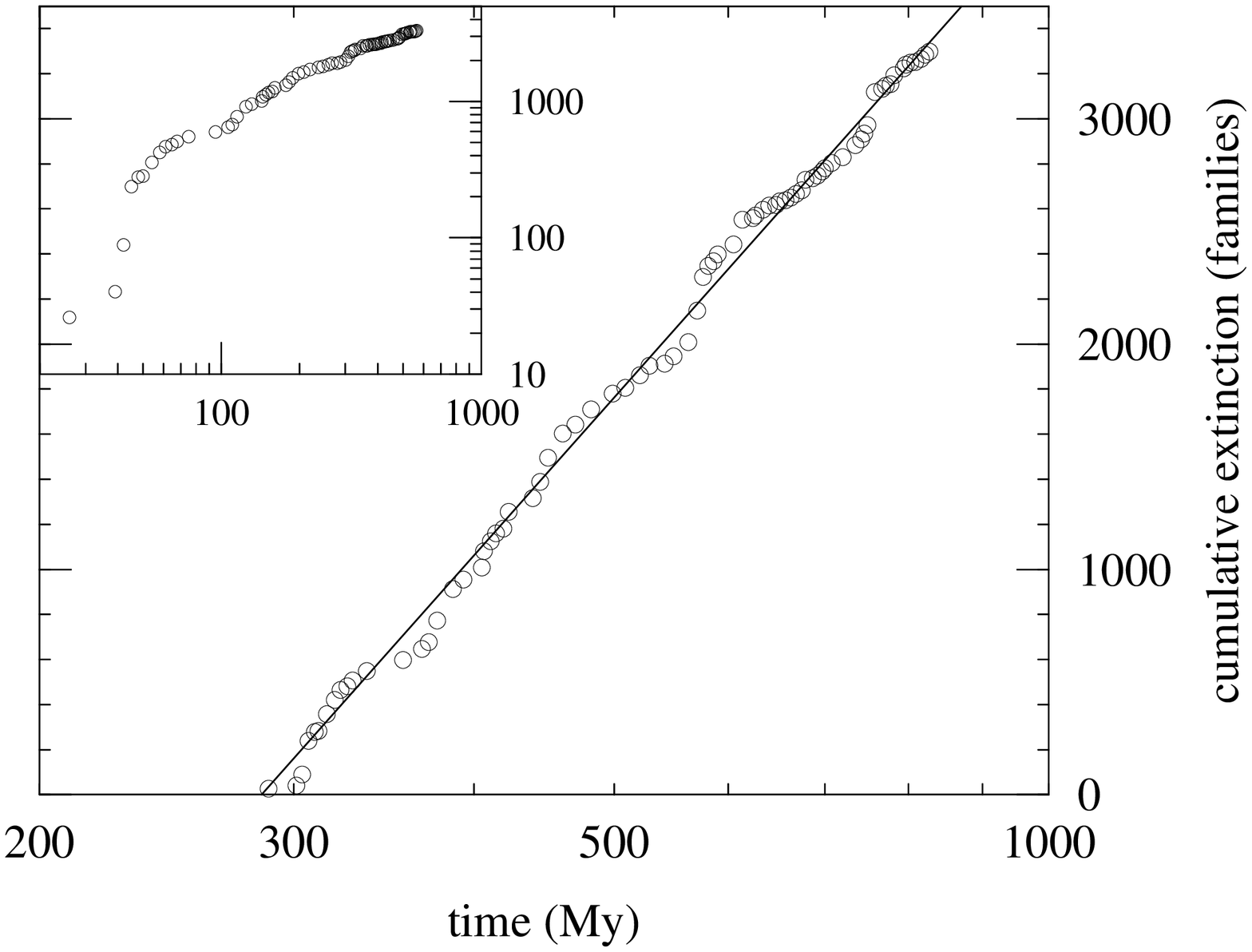}
\capt{Main figure: the cumulative extinction intensity as a function of
  time during the Phanerozoic on linear--log scales.  The straight line is
  the best logarithmic fit to the data.  Inset: the same data on log--log
  scales.}
\label{accum}
\end{figure}

In Figure~\fref{accum} we show the cumulative total extinction of marine
families, calculated from Sepkoski's data, as a function of time.  Note how
we have plotted the data.  The horizontal axis in the figure is logarithmic
and the vertical one is linear.  Plotted in this way, the cumulative
extinction appears as a straight line.  This implies that $c(t)$ increases
logarithmically in time.  In mathematical terms,
\begin{equation}
c(t) = A + B\log(t-t_0),
\label{logfit}
\end{equation}
where the best fit has an $R^2$ of $0.996$ and is obtained by choosing
values for the constants of $A=-17700$, $B=3130$, and $t_0=-262$~My.  The
last of these is a curious result.  It is interesting to speculate
(although we will not do so here) why the best fit should be obtained by
measuring time from a point 260~My before the start of the Cambrian.

Sibani~\etal~(1995, 1998) proposed that the curve of cumulative extinction
increases as a power law in time, meaning that it should appear straight on
dual logarithmic scales.  We have replotted the data in this way in the
inset to Figure~\fref{accum} but, as we can see, the fit in this case is
considerably poorer than our logarithmic one.  Interestingly, Sibani~\etal\ 
were studying cumulative extinction metrics in the context of a
quantitative model of extinction decline based on the idea of increasing
mean species fitness.  As it turns out, the predictions of their model are
precisely in agreement with the form, Equation~\eref{logfit}, found here,
rather than with their power law form.

Combining Equations~\eref{cumulative} and~\eref{logfit} and
differentiating, we now deduce the following form for the actual
(non-cumulative) total extinction rate:
\begin{equation}
x(t) = {B\over t-t_0}.
\label{decay}
\end{equation}
In other words, our analysis reveals that the decline in extinction during
the Phanerozoic is well parameterized by a function which falls off as the
reciprocal of time.  This functional form is shown as the dotted line in
Figure~\fref{decline}.  Similar results apply for other extinction metrics
as well.  The cumulated per-taxon extinction, for instance, also fits a
logarithmic growth law, with an $R^2$ of $0.991$.

In the next two sections we study the consequences of the decline in
extinction rate for the distributions of the sizes of extinction events and
for the power spectrum of extinction.

\section{Distribution of the sizes of extinction events}
One quantity which has received a lot of attention from the proponents of
self-organized critical theories of evolution is the distribution of the
sizes of extinction events in the fossil record.  In fact, the temporal
resolution of the fossil record is insufficient to distinguish individual
extinction events, and so attention has usually focussed on the number of
taxa becoming extinct per stratigraphic stage.  As Newman~(1996) has
pointed out, it is not clear that the distribution of this quantity obeys a
power law, but the data are compatible with the power-law form, with an
exponent of $-2.0\pm0.2$.  We now show that this is exactly what one should
expect from an extinction profile which decays over time according to
Equation~\eref{decay}.  Sibani~\etal~(1995) in their original paper
proposing the cumulative extinction measure demonstrated that given a
declining extinction rate one can extract an estimate of the expected
distribution of number of taxa becoming extinct per unit time.  Here we use
their methods to calculate this distribution for our data.

The number $\d t$ of unit time intervals in which the extinction rate has a
value in some range from $x$ to $x+\d x$ is proportional to the derivative
\begin{equation}
{\d t\over\d x} = -{(t-t_0)^2\over B} = -{B\over x^2},
\end{equation}
where we have made use of Equation~\eref{decay} twice.  In other words, the
probability $P(x)$ of a number $x$ of families becoming extinct in any
randomly-chosen stage should vary as
\begin{equation}
P(x) \sim x^{-2}.
\end{equation}
This is precisely the power-law form observed in the fossil data, with an
exponent $-2$ which agrees exactly with the fossil record.  Given the form
of the decline in extinction intensity therefore, no further assumptions
are needed to explain this power-law form.  In particular, it is not
necessary to assume a self-organized critical dynamics.

The argument above is not perfect.  In particular, it neglects the
variation in stage lengths, which could contribute significantly to
variation in the number of taxa becoming extinct per stage.  It also
assumes Equation~\eref{decay} to be an exact representation of the decline
in extinction rates, whereas it is of course only an approximation, since
the real extinction rate contains fluctuations about the average form, of
which the most important are the large mass extinction events.
Nonetheless, the crucial point is that the distribution of the sizes of
extinction events need not have anything to do with the proposed critical
processes.  A declining extinction rate is all that is needed to explain
the observed data.

\section{The power spectrum of extinction}
Sol\'e~\etal~(1997) have analysed the Phanerozoic extinction record using
Fourier transform techniques and proposed that the power spectrum $P(f)$ of
extinction falls off with frequency $f$ according, once more, to a power
law:
\begin{equation}
P(f) \sim f^{-\beta},
\end{equation}
with values of the exponent $\beta$ in the vicinity of 1.  They suggested
that this too could be a sign of critical dynamics in the processes giving
rise to extinction (although they also conceded that other explanations are
possible).  In this section we demonstrate that a mean extinction rate
which declines according to Equation~\eref{decay} is also sufficient to
explain the observed power spectrum, without resorting to critical
theories.

\begin{figure}[t]
\normalfigure{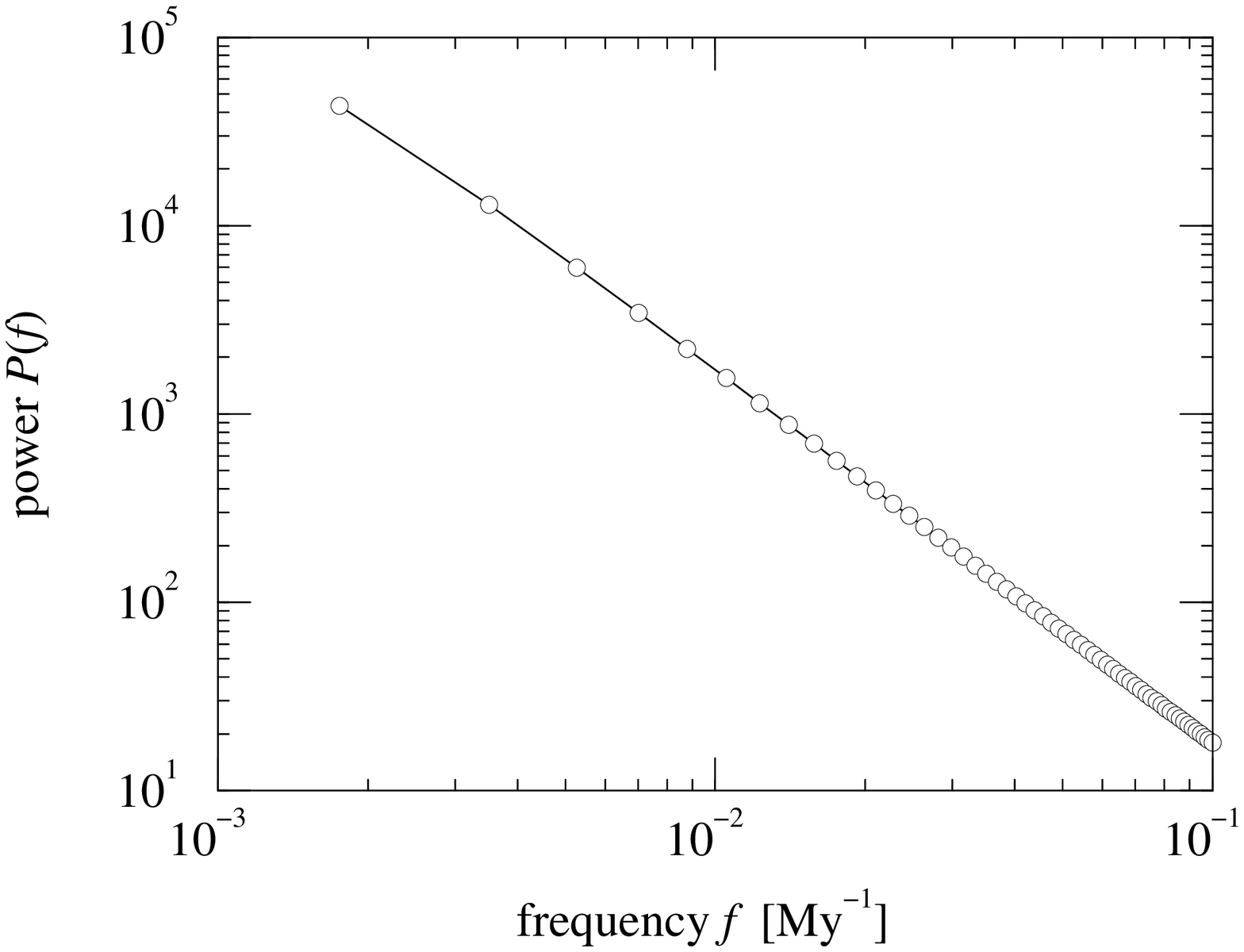}
\capt{The power spectrum for an extinction profile declining in accordance
  with Equation~\eref{decay}, calculated using Equation~(\ref{pf}).  The
  units on the vertical axis are arbitrary.}
\label{spectrum}
\end{figure}

The power spectrum for an extinction rate declining as $(t-t_0)^{-1}$ is
given by
\begin{equation}
P(f) = \left|\int_0^{t_1} (t-t_0)^{-1}\, {\rm e}^{{\rm i}2\pi\! ft} {\rm \d}t
       \right|^2,
\label{pf}
\end{equation}
where we measure time again from the beginning of the Phanerozoic and $t_1$
is the latest time for which we have data, in this case the end of the
Pleistocene.  This integral is unfortunately not analytically tractable,
but it is trivial to perform numerically.  We have done this and the
results are shown in Figure~\fref{spectrum} for the value of $t_0$
extracted from our fit to the fossil data.  As the figure shows, the
results follow a power-law form (i.e.,~a straight line on the logarithmic
scales employed in the figure) closely over the range of frequencies of
interest, with an exponent $\beta=1.96\pm0.01$.  Although, as we mentioned
above, Sol\'e~\etal\ found values of $\beta$ in the region of 1, in our own
calculations we have found values closer to $2$ (Newman and Eble~1998), so
the form found here is perfectly plausible.

It is worth mentioning that this is not the only alternative explanation
that has been proposed for the form of the fossil power spectrum.  In
another paper we discuss this issue in more detail and propose a different
interpretation of the results (Newman and Eble~1998).

\section{Conclusions}
In this paper we have studied the cumulative total extinction for fossil
families during the Phanerozoic and shown that it can be well approximated
by a logarithmic growth law.  This in turn implies that the extinction
intensity itself falls off as the reciprocal of time.  We have made use of
this observation to demonstrate possible explanations of two previously
published results: the power-law form of the distribution of the sizes of
extinction events, and the power-law form of the power spectrum of the
fossil extinction record.  In both cases we show that the decline in
extinction intensity is all that is necessary to explain the observed
results; we do not need to appeal to an underlying self-organized critical
dynamics as other authors have done.

The results presented in this paper are only one step in an argument.  We
have shown that a declining extinction rate can account for a number of
observations made by other authors.  The other step in the argument is to
explain what gives rise to the decline.  This is a question of great
interest and one about which arguments will no doubt continue for many
years to come.

\section{Acknowledgements}
The authors would like to thank Jack Sepkoski for providing us with the
fossil data used in this study, and Doug Erwin, David Raup, Jack Sepkoski,
Paolo Sibani and Ricard Sol\'e for illuminating discussions and comments.
This work was funded in part by the Smithsonian Institution and by the
Santa Fe Institute and DARPA under grant number ONR N00014--95--1--0975.

\def\refer#1#2#3#4#5#6{\item{\frenchspacing\sc#1}\hspace{4pt}
                       #2\hspace{8pt}#3  {#4} {\bf#5}, #6.}
\def\bookref#1#2#3#4{\item{\frenchspacing\sc#1}\hspace{4pt}
                     #2\hspace{8pt}{\it#3}  #4.}

\newpage

\section*{References}
\baselineskip=15pt

\begin{list}{}{\leftmargin=2em \itemindent=-\leftmargin%
\itemsep=3pt \parsep=0pt \small}

\item {\sc Bak, P. and Paczuski, M.}\ \ 1996\ \ Mass extinctions vs.\ 
  uniformitarianism in biological evolution.  In {\it Physics of
  Biological Systems,} Springer--Verlag, Berlin.

\refer{Bak, P. \& Sneppen, K.}{1993}{Punctuated equilibrium and
  criticality in a simple model of
  evolution.}{\it Phys. Rev. Lett.}{71}{4083--4086}

\refer{Bak, P., Tang, C. \& Wiesenfeld, K.}{1987}{Self-organized
  criticality: An explanation of $1/f$ noise.}{\it
  Phys. Rev. Lett.}{59}{381--384}

\bookref{Benton, M. J.}{1993}{The Fossil Record 2.}{Chapman and Hall
  (London)}

\refer{Benton, M. J.}{1995}{Diversification and extinction in the
  history of life.}{\it Science\/}{268}{52--58}

\refer{Bowring, S. A., Grotzinger, J. P., Isachsen, C. E., Knoll, A. H.,
  Pele\-chaty, S. M. \& Kolosov, P.}{1993}{Calibrating rates of early
  Cambrian evolution.}{\it Science\/}{261}{1293--1298}
  
\refer{Flessa, K. W. \& Jablonski, D.}{1985}{Declining Phanerozoic
  background extinction rates: Effect of taxonomic structure?}{\it
  Nature\/}{313}{216--218}

\refer{Gilinsky, N. L.}{1994}{Volatility and the Phanerozoic decline of
background extinction intensity.}{\it Paleobiology\/}{20}{445--458}

\refer{Gilinsky, N. L. and Bambach, R. K.}{1987}{Asymmetrical patterns of
  origination and extinction in higher taxa.}{\it
  Paleobiology\/}{13}{427--445}

\bookref{Harland, W. B., Armstrong, R., Cox, V. A., Craig, L. E.,
  Smith, A. G. \& Smith, D. G.}{1990}{A Geologic Time Scale
  1989.}{Cambridge University Press (Cambridge)}

\bookref{Kauffman, S. A.}{1993}{Origins of Order: Self-Organization and
  Selection in Evolution.}{Oxford University Press (Oxford)}

\refer{Kauffman, S. A. \& Johnsen, S.}{1991}{Coevolution to the edge of
  chaos: Coupled fitness landscapes, poised states, and coevolutionary
  avalanches.}{\it J. Theor. Biol.}{149}{467--505}

\refer{Newman, M. E. J.}{1996}{Self-organized criticality, evolution
  and the fossil extinction record.}{\it Proc. R. Soc. B\/}{263}{1605--1610}

\item {\sc Newman, M. E. J. and Eble, G. J.}\ \ 1998\ \ Power spectra of
  extinction in the fossil record.  Submitted to {\it Proc.\ R.\ Soc.\ 
  Lond.}

\refer{Pease, C. M.}{1992}{On the declining extinction and origination rates
  of fossil taxa.}{\it Paleobiology\/}{18}{89--92}

\item {\sc Raup, D. M.}\ \ 1988\ \ Testing the fossil record for
  evolutionary progress.  In {\it Evolutionary Progress,} Nitecki,
  M. (ed.), University of Chicago Press, Chicago.

\refer{Raup, D. M. and Sepkoski, J. J., Jr.}{1982}{Mass extinctions in the
  marine fossil record.}{\it Science\/}{215}{1501--1503}

\refer{Sepkoski, J. J., Jr.}{1984}{A kinetic model of Phanerozoic taxonomic
  diversity III.  Post-Paleozoic families and mass extinctions.}{\it
  Paleobiology\/}{10}{246--267}

\refer{Sepkoski, J. J., Jr.}{1991}{A model of onshore--offshore change in
  faunal diversity.}{\it Paleobiology\/}{17}{58--77}

\item {\sc Sepkoski, J. J., Jr.}\ \ 1992\ \ A compendium of fossil marine
  animal families, 2nd edition.  {\it Milwaukee Public Museum Contributions
    in Biology and Geology\/} {\bf83}.

\refer{Sepkoski, J. J., Jr.}{1993}{Ten years in the library: New data
  confirm paleontological patterns.}{\it Paleobiology\/}{19}{43--51}

\item {\sc Sepkoski, J. J., Jr.}\ \ 1996\ \ Patterns of Phanerozoic
  extinction: A perspective from global databases.  In {\it Global events
  and event stratigraphy,} Walliser, O. H. (ed.), Springer--Verlag, Berlin.

\refer{Sibani, P., Schmidt, M. R. and Alstr\o{}m, P.}{1995}{Fitness
  optimization and decay of extinction rate through biological
  evolution.}{\sl Phys. Rev. Lett.}{75}{2055--2058}

\refer{Sibani, P., Schmidt, M. R. and Alstr\o{}m, P.}{1998}{Evolution and
  extinction dynamics in rugged fitness landscapes.}{\sl
  Int. J. Mod. Phys. B\/}{12}{361--391}

\refer{Sneppen, K., Bak, P., Flyvbjerg, H. \& Jensen,
  M. H.}{1995}{Evolution as a self-organized critical
  phenomenon.}{\it Proc. Natl. Acad. Sci.}{92}{5209--5213}

\refer{Sol\'e, R. V. \& Bascompte, J.}{1996}{Are critical phenomena
  relevant to large-scale evolution?}{\it Proc. R. Soc. B\/}{263}{161--168}

\refer{Sol\'e, R. V., Manrubia, S. C., Benton, M. \& Bak,
  P.}{1997}{Self-similarity of extinction statistics in the fossil
  record.}{\it Nature\/}{388}{764--767}

\refer{Van Valen, L. M.}{1984}{A resetting of Phanerozoic community
  evolution.}{\it Nature\/}{307}{50--52}

\end{list}

\end{document}